\documentclass[twocolumn,twoside,slac_two]{revtex4}
\usepackage{hyperref}
\usepackage{amssymb}
\usepackage{amsmath}
\usepackage[pdftex]{graphicx}
\usepackage{fancyhdr}
\begin{document}

\title{\textit{Fermi} observations of TeV AGN}
\author{Stephen Fegan, David Sanchez}
\affiliation{\,LLR/Ecole Polytechnique/CNRS/IN2P3}
\author{On behalf of the \textit{Fermi}-LAT collaboration}

\begin{abstract}
We report on observations of TeV-selected AGN made during the first
5.5 months of observations with the Large Area Telescope (LAT)
on-board the \textit{Fermi} Gamma-ray Space Telescope (\textit{Fermi}). In total, 28 TeV
AGN were selected for study. The \textit{Fermi} observations show clear
detections of 21 of these TeV-selected objects. Most can be described
with a power law of spectral index harder than 2, with a spectral
break generally required to accommodate the TeV measurements. Evidence
for systematic evolution of the gamma-ray spectrum with redshift is
presented and discussed in the context of the EBL.
\end{abstract}

\maketitle

\section{Introduction}

At energies above approximately 100GeV (the TeV energy regime),
ground-based gamma-ray observatories have detected 96 sources over the
past two decades.  The pace of discovery in this energy regime has
been particularly high since the inception of the latest generation of
instruments: H.E.S.S., CANGAROO, MAGIC and VERITAS. The majority of
the TeV sources are galactic, however 30 extragalactic sources have
also been detected, of which 28 correspond to Active Galactic Nuclei
(AGN), the other two being starburst galaxies. Most (25) of these TeV
AGN are blazars.

\begin{figure}
\includegraphics[width=60mm]{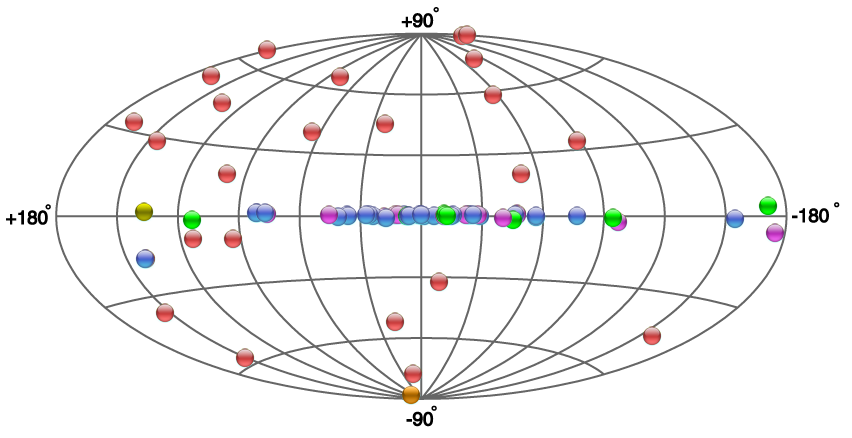}
\caption{\label{FIG::TEVCAT} TeV sources, from TeVCat
(\url{http://tevcat.uchicago.edu/).}}
\end{figure}

The Large Area Telescope is a pair-conversion telescope on the \textit{Fermi}
Gamma-ray Space Telescope (formerly GLAST), launched in June 2008. The
\textit{Fermi}-LAT instrument detects gamma rays with energies between 20 MeV
and $>$300 GeV (the GeV energy regime). In this poster we present the
results of \textit{Fermi}-LAT observations of the known TeV blazars. The
motivation for this study is two-fold: (i) to present as complete a
picture of the high-energy emission as possible by combining the GeV
and TeV results on these objects, and (ii) to help guide future TeV
observations. For a selection of these GeV--TeV objects we present the
GeV spectrum from \textit{Fermi} and extrapolate it to TeV energies assuming
absorption on the EBL, and compare these extrapolations to archival
TeV measurements. Finally, we study the evolution of the spectrum of
these objects as a function of redshift.

\section{Results summary -- Detected GeV--TeV AGN}

Table~\ref{TAB::RESULTS} presents the results of 5.5 months of
observation of the TeV AGN with the \textit{Fermi} LAT \citep[for
detailed discussion of analysis and results and a comprehensive list
of references to TeV data see][]{REF::TEVAGN}. Of the 28 objects
selected for observation, a total of 21 were detected with TS$>$25
(approximately $5^\circ$). This degree of connection between the TeV
blazars and the GeV regime was not found by EGRET and the previous
generation of TeV instruments, and is evident now only as a result of
the improved sensitivity and greater overlap between the effective
energy ranges of \textit{Fermi} and the current generation of TeV
instruments.

\begin{table*}[p]
\centering
\caption{\label{TAB::RESULTS} TeV AGN detected by the \textit{Fermi}/LAT 
in 5.5 months of observation}
\includegraphics[width=0.8\textwidth]{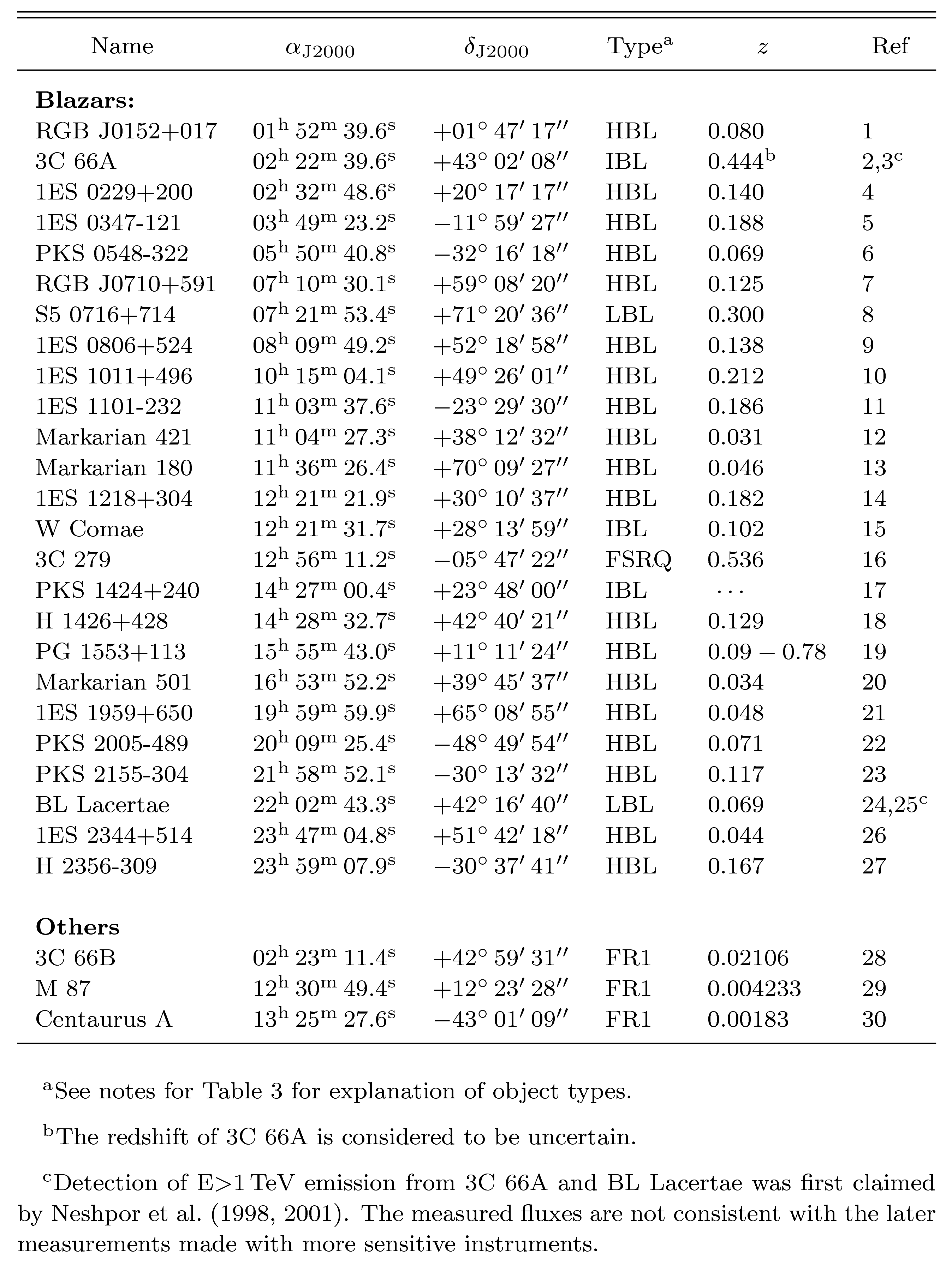}
\end{table*}

The majority of the TeV blazars detected by \textit{Fermi} have a photon index
$\Gamma\le2$ in the GeV regime, the median index is $\Gamma=1.9$. In
contrast, the populations of 42 BL Lacs and 57 FSRQs from the LBAS
sample have median indexes of $\Gamma=2.0$ and $\Gamma=2.4$
respectively. The TeV blazars are amongst the hardest extragalactic
objects detected by \textit{Fermi}. For many of the sources, especially those
with harder spectra, no evidence for curvature is seen in the LAT
energy range. Furthermore, many sources did not show evidence of
significant variability over the period of the study.

\section{Discussion of selected AGN}

\begin{figure}
\includegraphics[width=60mm]{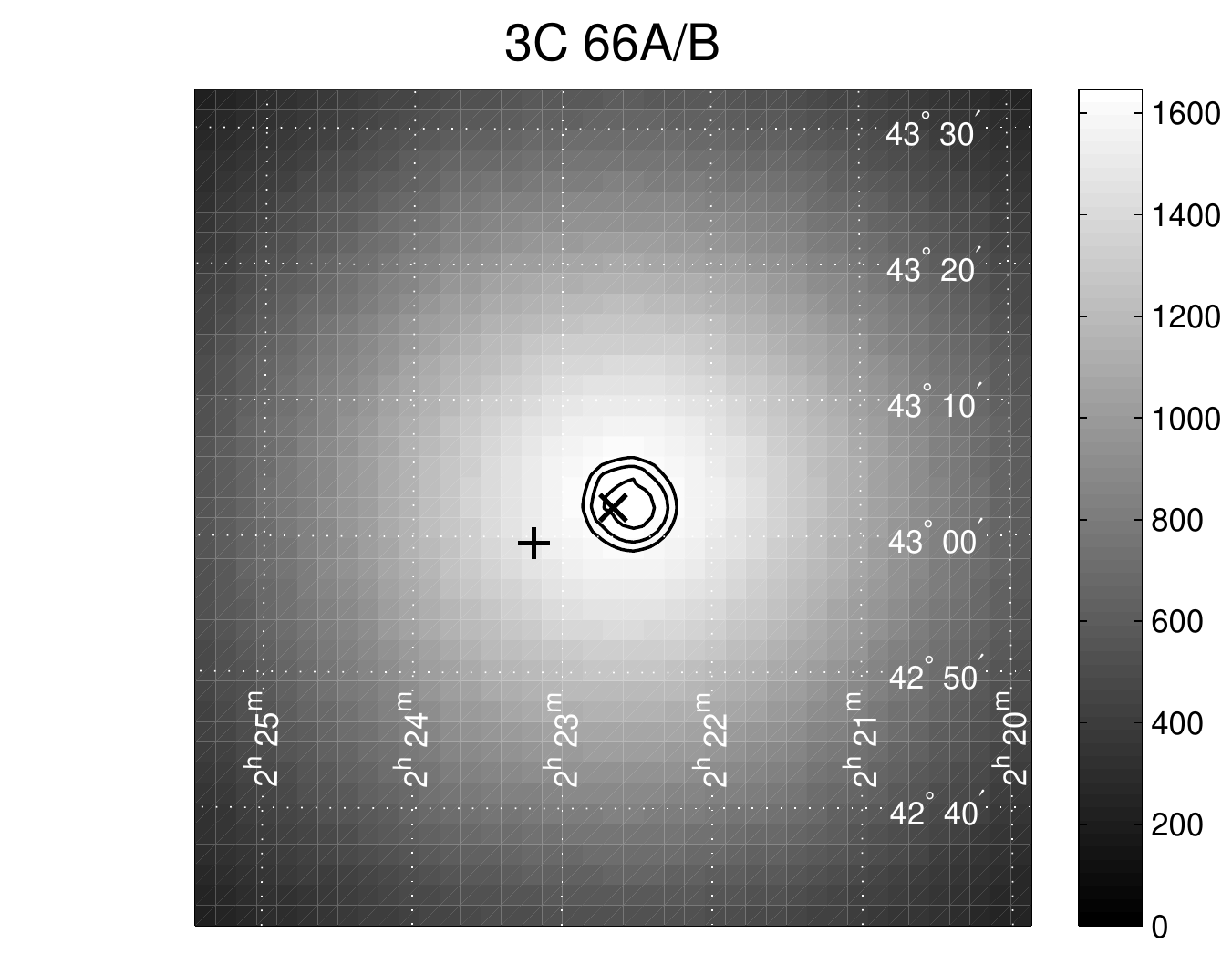}
\caption{\label{FIG::3C66A_TS} TS map of the 3C~66A/B region.}
\end{figure}

\begin{figure}
\includegraphics[width=60mm]{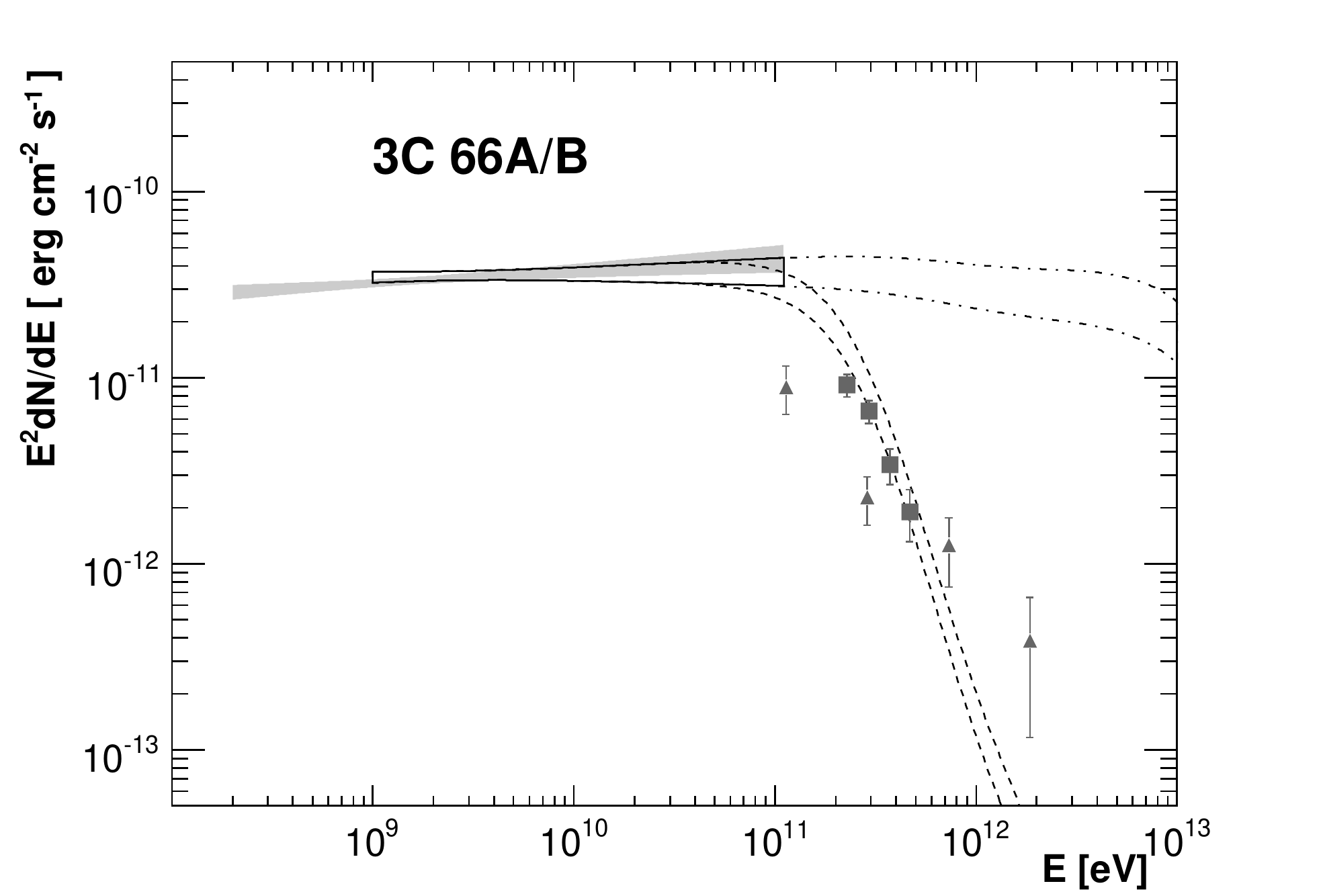}
\caption{\label{FIG::3C66A_SPEC} Spectrum for 3C~66A/B region.}
\end{figure}

\textbf{3C 66A/B:} TeV emission from this region detected by VERITAS
(3C~66A, HBL, $z=0.444$) and MAGIC (3C~66B, RG, $z=0.0211$).  \textit{Fermi}
detects hard GeV emission, coincident with 3C~66A
(Figures~\ref{FIG::3C66A_TS} and \ref{FIG::3C66A_SPEC}). An
extrapolation of the GeV spectrum to TeV energies is in better
agreement with the TeV data measurements assuming $z=0.444$ than
$z=0.0221$.

\begin{figure}
\includegraphics[width=60mm]{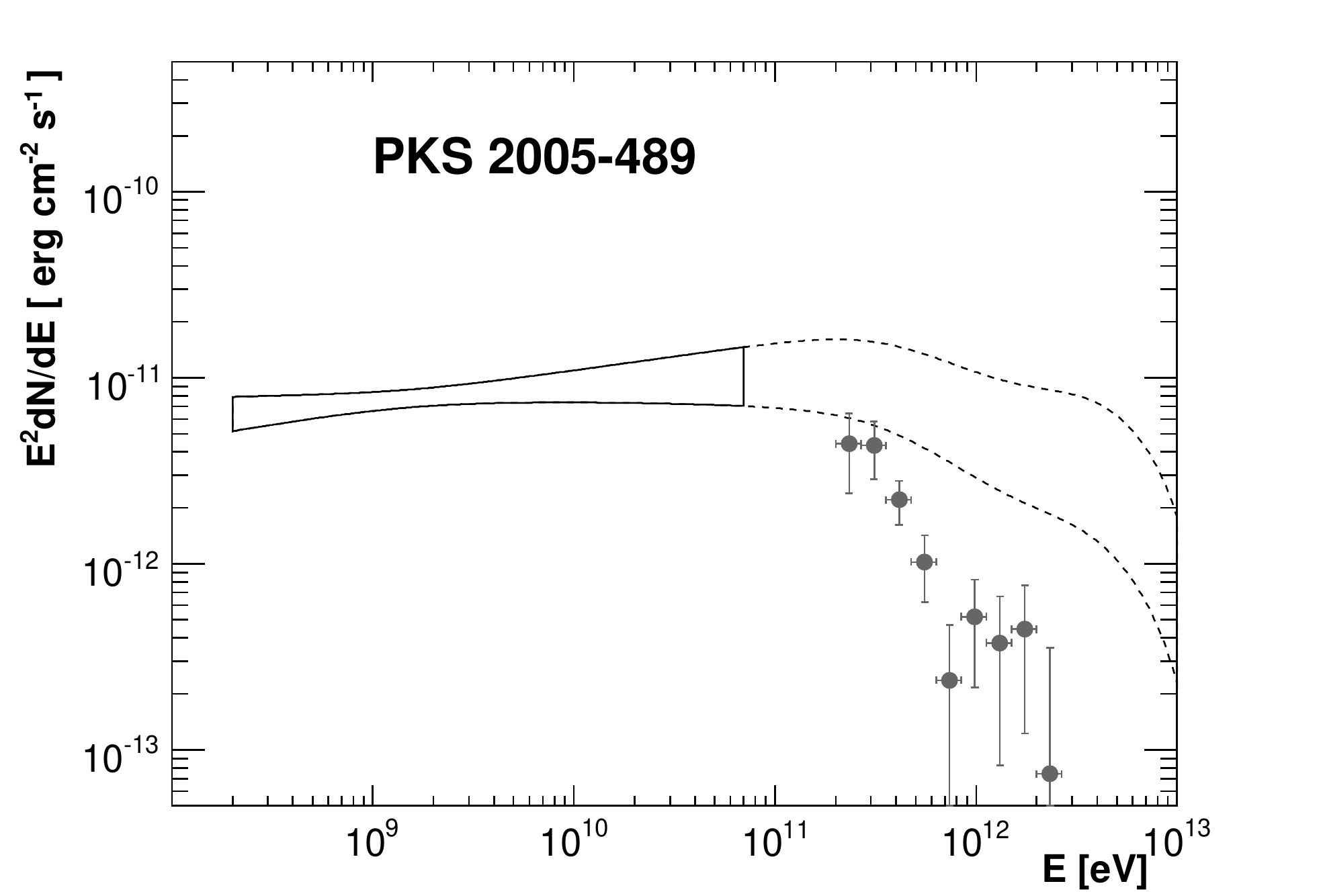}
\caption{\label{FIG::pks2005} Spectrum for PKS~2005-489.}
\end{figure}

\textbf{PKS 2005-489:} A H.E.S.S.-detected HBL with no evidence of TeV
variability. \textit{Fermi} detects hard steady emission from this source
(Figure~\ref{FIG::pks2005}). An extrapolation of the GeV
spectrum to TeV energies over- predicts the TeV spectrum, suggesting
the presence of intrinsic curvature in the spectrum of PKS~2005-489.

\begin{figure}
\includegraphics[width=60mm]{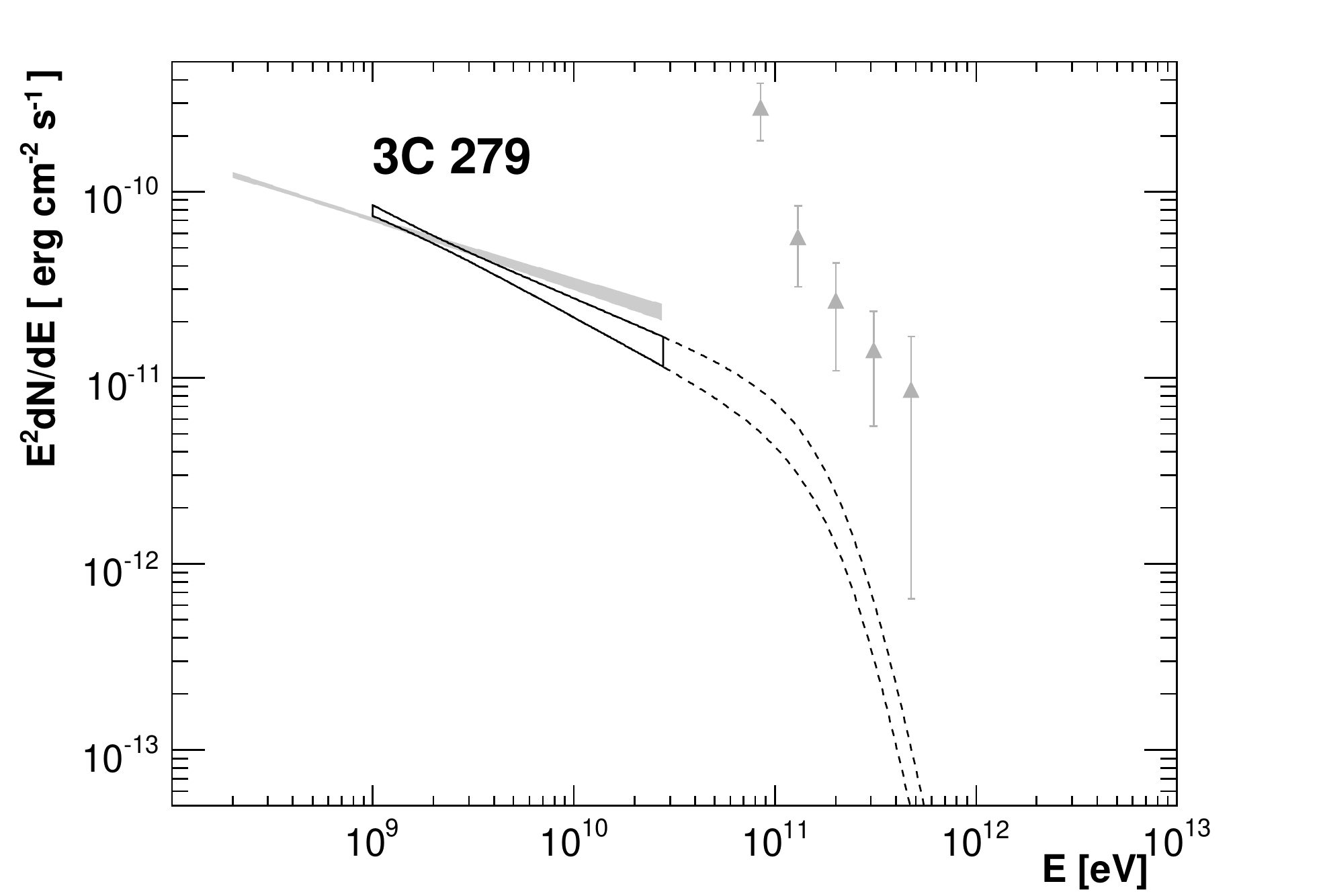}
\caption{\label{FIG::3c279} Spectrum for 3C~279.}
\end{figure}

\textbf{3C 279:} Detected by MAGIC during a flaring episode, 3C~279 is
the most distant TeV source (with a known redshift) detected to date.
The \textit{Fermi} spectrum is relatively soft, and shows clear evidence for
curvature (Figure~\ref{FIG::3c279}). During the period
of the the study the flux from 3C~279 increased by a factor of $\sim5$
in a flare that lasted $\sim50$ days.  The spectra for the flaring and
non-flaring emission is shown. An extrapolation of both to TeV
energies under-predicts the flux measured by MAGIC, showing that it
must correspond to a extreme flaring state.

\section{Evolution of spectra with redshift}

\begin{figure}
\centerline{\includegraphics[width=60mm]{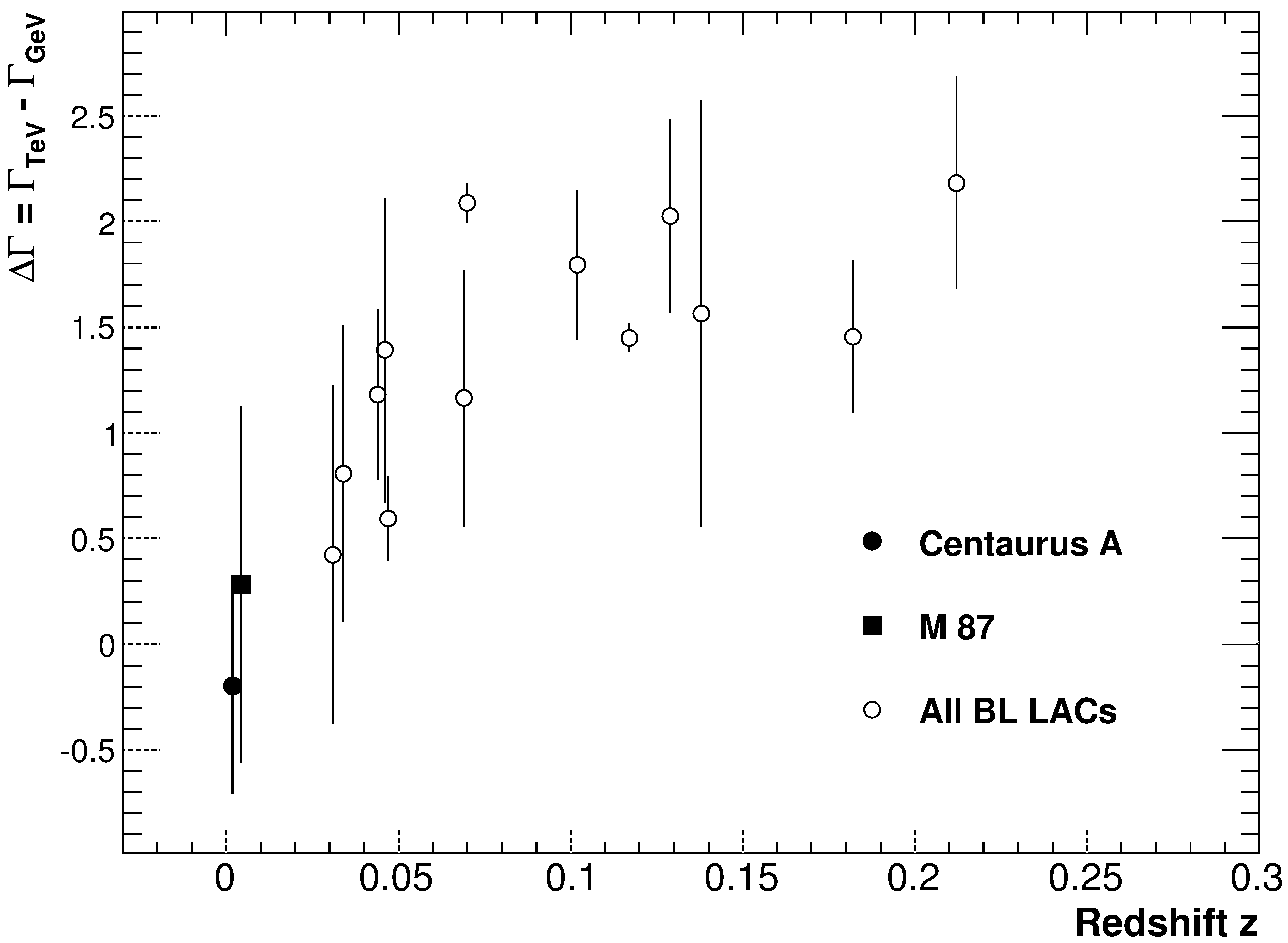}}
\caption{\label{FIG::DGAMMA} Difference in TeV and GeV spectral indices for GeV--TeV AGN.}
\end{figure}

In the LBAS study \citep{REF::LBAS}, no significant relation between
the GeV photon index and redshift was found for either the sample of
BL Lacs or FSRQs. The GeV--TeV sources provide a population in which
the effects of spectral evolution with redshift can be studied across
a much wider energy range than LBAS.  The presence of a
redshift-dependent spectral break in these sources could be indicative
of the effects of absorption on the EBL, and provide experimental
evidence for this absorption in a manner independent of any specific
EBL-density model.  The difference in the TeV and GeV spectral indices
for 15 of the GeV--TeV sources is shown in Figure~\ref{FIG::DGAMMA}.
It is evident that the difference between the GeV and TeV spectral
indices increases with redshift. At low redshifts the radio galaxies
M~87 and Cen~A have $\Delta\Gamma\sim0$, as do the near-by BL Lacs. At
$z>0.1$, all of the BL Lacs are consistent with $\Delta\Gamma\ge1.5$.

\section{Conclusions}

In 5.5 months of observation the \textit{Fermi} LAT has detected GeV emission
from 21 TeV-selected AGN (and from 17 previously observed by TeV
groups for which upper limits have been published). Many exhibit an
increasing spectrum ($\Gamma<2$) in the GeV range confirming the
presence of a high energy peak in $\nu F_\nu$ representation. The intrinsic
spectrum for some of the TeV sources can be well described by a single
power-law across the energy range spanned by the \textit{Fermi} LAT and the TeV
observatories, with any breaks in the measured gamma-ray spectra
between the two regimes being consistent with the effects of
absorption with a model of minimal EBL density. Redshift-dependent
evolution is detected in the spectra of objects detected at GeV and
TeV energies. The most reasonable explanation for this is absorption
on the EBL.

\bigskip 
\begin{acknowledgments}

The \textit{Fermi}-LAT Collaboration acknowledges the generous support of a
number of agencies and institutes that have supported the \textit{Fermi}-LAT
Collaboration. These include the National Aeronautics and Space
Administration and the Department of Energy in the United States, the
Commissariat \`a l'Energie Atomique and the Centre National de la
Recherche Scientifique / Institut National de Physique Nucl\'eaire et
de Physique des Particules in France, the Agenzia Spaziale Italiana
and the Istituto Nazionale di Fisica Nucleare in Italy, the Ministry
of Education, Culture, Sports, Science and Technology (MEXT), High
Energy Accelerator Research Organization (KEK) and Japan Aerospace
Exploration Agency (JAXA) in Japan, and the K.\ A.\ Wallenberg
Foundation, the Swedish Research Council and the Swedish National
Space Board in Sweden.

Additional support for science analysis during the operations phase is
gratefully acknowledged from the Istituto Nazionale di Astrofisica in
Italy.

This research has made use of NASA's Astrophysics Data System
Bibliographic Services, the NASA/IPAC Extragalactic Database, operated
by JPL, Caltech, under contract from NASA, and the SIMBAD database,
operated at CDS, Strasbourg, France.
\end{acknowledgments}

\bigskip 

\end{document}